\documentclass[aps,twocolumn,amsmath,amssymb,superscriptaddress]{revtex4-1}
\usepackage{graphicx}
\usepackage{dcolumn}
\usepackage{graphics}
\usepackage{bbm}
\usepackage{amsmath,amsfonts,amssymb}
\usepackage{wrapfig}
\begin{document}
\title{Dispersive Readout of a Few-Electron Double Quantum Dot with Fast rf Gate-Sensors}
\author{J. I. Colless}
\affiliation{ARC Centre of Excellence for Engineered Quantum Systems, School of Physics, The University of Sydney, Sydney, NSW 2006, Australia.}
\author{A. C. Mahoney}
\affiliation{ARC Centre of Excellence for Engineered Quantum Systems, School of Physics, The University of Sydney, Sydney, NSW 2006, Australia.}
\author{J. M. Hornibrook}
\affiliation{ARC Centre of Excellence for Engineered Quantum Systems, School of Physics, The University of Sydney, Sydney, NSW 2006, Australia.}
\author{A. C. Doherty}
\affiliation{ARC Centre of Excellence for Engineered Quantum Systems, School of Physics, The University of Sydney, Sydney, NSW 2006, Australia.}
\author{D. J. Reilly$^\dagger$}
\affiliation{ARC Centre of Excellence for Engineered Quantum Systems, School of Physics, The University of Sydney, Sydney, NSW 2006, Australia.}
\author{H. Lu}
\affiliation{Materials Department, University of California, Santa Barbara, California 93106, USA.}
\author{A. C. Gossard}
\affiliation{Materials Department, University of California, Santa Barbara, California 93106, USA.}

\begin{abstract}
We report the dispersive charge-state readout of a double quantum dot in the few-electron regime using the $in$ $situ$ gate electrodes as sensitive detectors. We benchmark this gate-sensing technique against the well established quantum point contact (QPC) charge detector and find comparable performance with a bandwidth of $\sim$ 10 MHz and an equivalent charge sensitivity of $\sim$ 6.3 $\times$ $10^{-3}$ e/$\sqrt{\mathrm{Hz}}$. Dispersive gate-sensing alleviates the burden of separate charge detectors for quantum dot systems and promises to enable readout of qubits in scaled-up arrays.
\end{abstract}
\maketitle

Non-invasive charge detection has emerged as an important tool for uncovering new physics in nanoscale devices at the single-electron level and allows readout of spin qubits in a variety of material systems \cite{Field,Sprinzak,Rimberg,Elzerman:803014,DiCarlo,Petta:2005kn,Amasha:2008ky,Mingyun,Maune}.  For quantum dots defined electrostatically by the selective depletion of a two dimensional electron gas (2DEG), the conductance of a proximal quantum point contact (QPC) \cite{Elzerman:803014,DiCarlo,Petta:2005kn,Amasha:2008ky,Maune} or single electron transistor (SET) \cite{Rimberg,Mingyun} can be used to detect the charge configuration in a regime where direct transport is not possible.  This method can, in principle, reach quantum mechanical limits for sensitivity \cite{Devoret_schoel} and has enabled the detection of single electron spin-states \cite{Elzerman:803014,Amasha:2008ky,Barthel:2009hx} with a 98\% readout fidelity in a single-shot \cite{Yacoby2qubit}.

An alternate approach to charge-state detection, long used in the context of single electron spectroscopy \cite{Ashoori}, is based on the dispersive signal from shifts in the quantum capacitance \cite{McEuen,Fujisawa} when electrons undergo tunnelling. Similar dispersive interactions are now the basis for  readout in a variety of quantum systems including atoms in an optical resonator \cite{Heinzen}, superconducting qubits \cite{Duty,Roschier,Wallraff} and nanomechanical devices \cite{Roukes}. 

In this Letter we report dispersive readout of quantum dot devices using the standard, $in$ $situ$ gate electrodes coupled to lumped-element resonators as high-bandwidth, sensitive charge-transition sensors. We demonstrate the sensitivity of this gate-sensor in the few-electron regime, where these devices are commonly operated as charge or spin qubits \cite{Hanson:2007eg} and benchmark its performance against the well established QPC charge sensor. We find that because the quantum capacitance is sufficiently large in these devices, gate-sensors have similar sensitivity to QPC sensors. In addition, we show that gate-sensors can operate at elevated temperatures in comparison to QPCs.  

\begin{figure}
\includegraphics[scale=0.6]{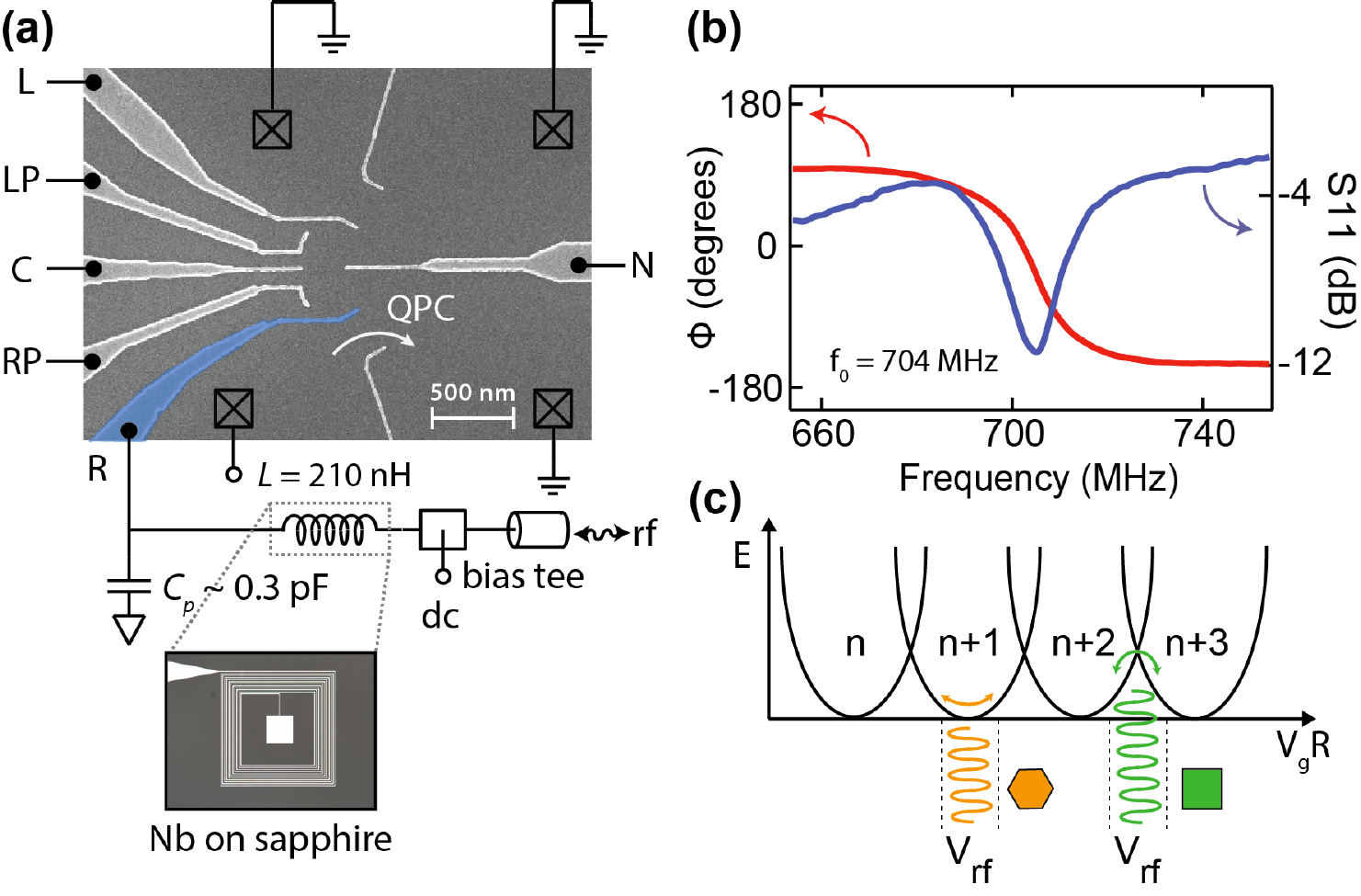}
\caption{\label{fig:photos}\textbf{(a)} Micrograph of a similar device to the one measured and circuit schematic. One of the $in$ $situ$ dot-defining gates (blue) is coupled via a bondwire to an off-chip Nb/Al$_2$O$_3$ superconducting lumped-element resonator to enable dispersive readout. \textbf{(b)} Amplitude S$_{11}$ (blue) and phase response (red) of the resonator. \textbf{(c)} Illustration of the charging energy spectrum for a quantum dot.  The resonant rf gate voltage $V_{rf}$ induces tunnelling at the charge degeneracy point (green oscillation) leading to a dispersive shift that is suppressed for configurations of stable charge (orange oscillation).}
\vspace{-0.5cm}
\end{figure} 

Previous investigations, in the context of circuit quantum electrodynamics (c-QED), have engineered a dispersive interaction between many-electron dots and superconducting coplanar waveguide resonators \cite{Frey,Toida,FreyPRB,PeterssonPetta}. Recently, the charge and spin configuration of double quantum dots has also been detected by dispersive changes in a radio frequency resonator coupled directly to the source or drain contacts of the device \cite{Petersson,Chorley,PeterssonPetta,Petta_wire}. The present work advances these previous studies by demonstrating that the gates, already in place to define the quantum dot system, can also act as fast and sensitive readout detectors in the single-electron regime. This is a surprising result, given the small capacitive coupling between the gate and dot, but lifts a barrier to qubit readout in large scaled-up quantum dot arrays by alleviating the need for many ohmic contacts, large on-chip distributed resonators, or proximal charge detectors. 

Our gate-sensor, shown in Fig. 1(a), comprises an off-chip superconducting Nb on Al$_2$O$_3$ spiral inductor ($L \sim$ 210 nH) in resonance with the distributed parasitic capacitance ($C \sim$ 0.3 pF) that includes a TiAu gate electrode used to define the quantum dots (resonance frequency $f_0$ = 1/$\sqrt{{\mathrm{L C}}}$ = 704 MHz, $Q$-factor $\sim$ 70). The dots are 110 nm below the surface of a GaAs/Al$_{0.3}$Ga$_{0.7}$As heterostructure (electron density 2.4 $\times$ 10$^{15}$ m$^{-2}$, mobility 44 m$^2$/V s) that is mounted on a high-frequency circuit board \cite{Colless} at the mixing chamber of a dilution refrigerator with base temperature $T$ $\sim$ 20 mK. The electron temperature $T_e$, determined by Coulomb blockade (CB) thermometry, is below 100 mK. The amplitude and phase response of the resonator is measured, following cryogenic amplification, using a vector network analyzer, as shown in Fig. 1(b). 
\begin{figure}
\includegraphics[scale=0.55]{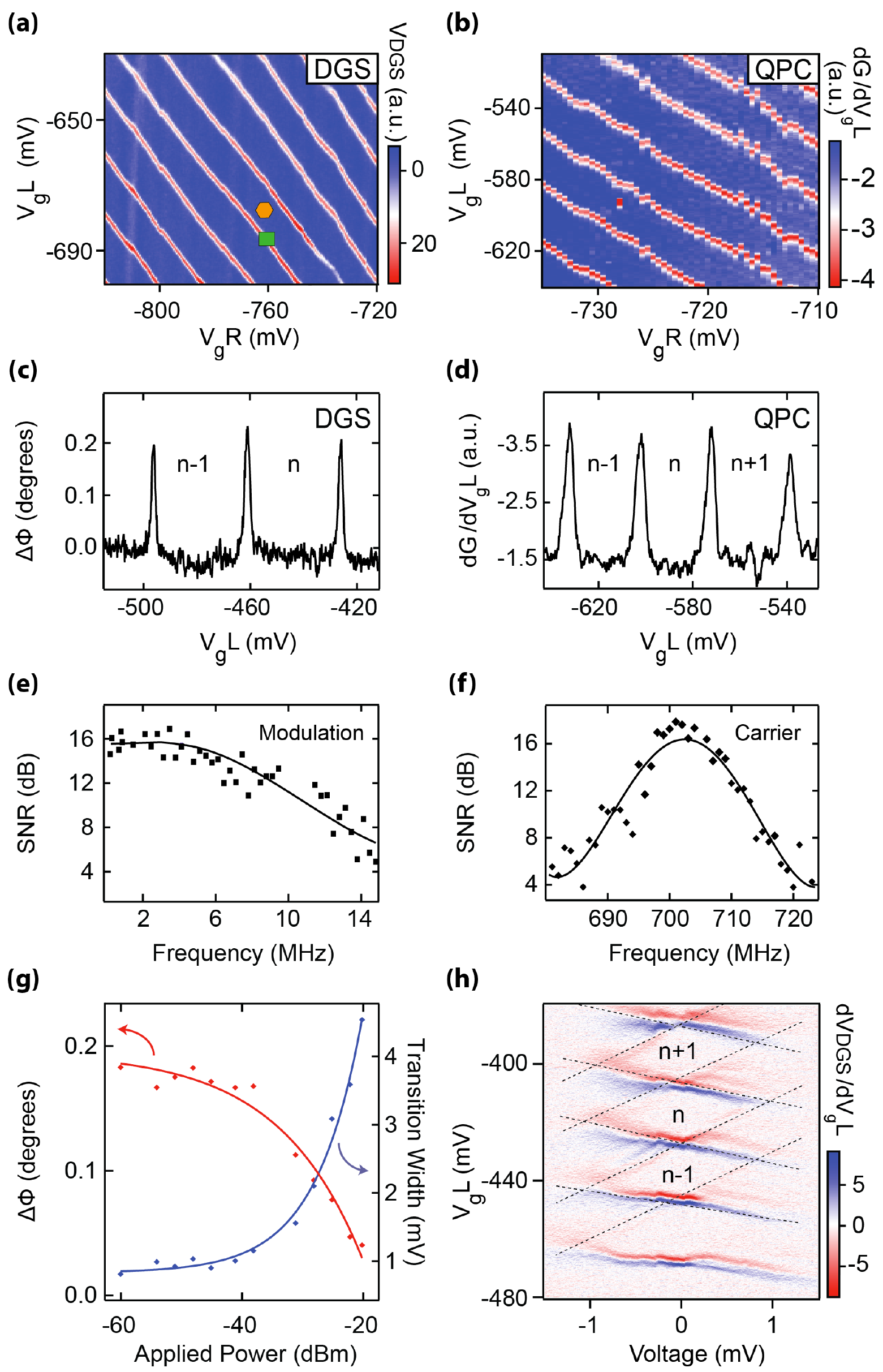}
\caption{\textbf{(a)} Dispersive signal from the gate-sensor showing transitions in electron number for a large single quantum dot. Green and orange symbols correspond to positions of symbols in Fig. 1(c). \textbf{(b)} Derivative of the QPC conductance signal with gate voltage $V_gL$  in a region of gate-space similar to (a). The slight shift in gate voltage and period of the oscillations in comparison to (a) is due to the presence of the QPC gate bias.  \textbf{(c)} Phase response of the gate-sensor showing peaks corresponding to single electron transitions. \textbf{(d)} Vertical slice through the conductance signal in (b), at $V_gR$ = -723 mV. \textbf{(e)} SNR of the gate-sensor as a function of the modulation frequency of a signal applied to a nearby gate. \textbf{(f)} SNR for the gate-sensor as a function of carrier frequency. \textbf{(g)} Width and height of the DGS response signal with power applied to the resonator (measured before attenuation). \textbf{(h)} Coulomb charging diamonds for the quantum dot, measured using the gate-sensor in a regime where direct transport is not possible. Colour scale is the derivative of the dispersive signal. Labels indicate number of electrons.}
\vspace{-0.5cm}
\label{fig2}
\end{figure}

Dispersive gate-sensors (DGS) detect charge-transitions (rather than absolute charge) by sensing small changes in the polarizability or quantum admittance \cite{FreyPRB} when an electron tunnels in response to the alternating rf gate voltage. Tunnelling modifies the resonator capacitance $C$ beyond the geometric contribution (at the position of green symbol in Fig. 1(c)) compared to the regime where tunnelling is suppressed (orange symbol in Fig. 1(c)). The response of the resonator $\Delta \phi$ is detected by fast sampling of the in-phase and quadrature components of the reflected rf to produce a baseband signal, $V_{DGS}$, proportional to the dispersive shift \cite{Reilly:2007ig}.
\begin{figure*}\centering
\includegraphics[scale=0.6]{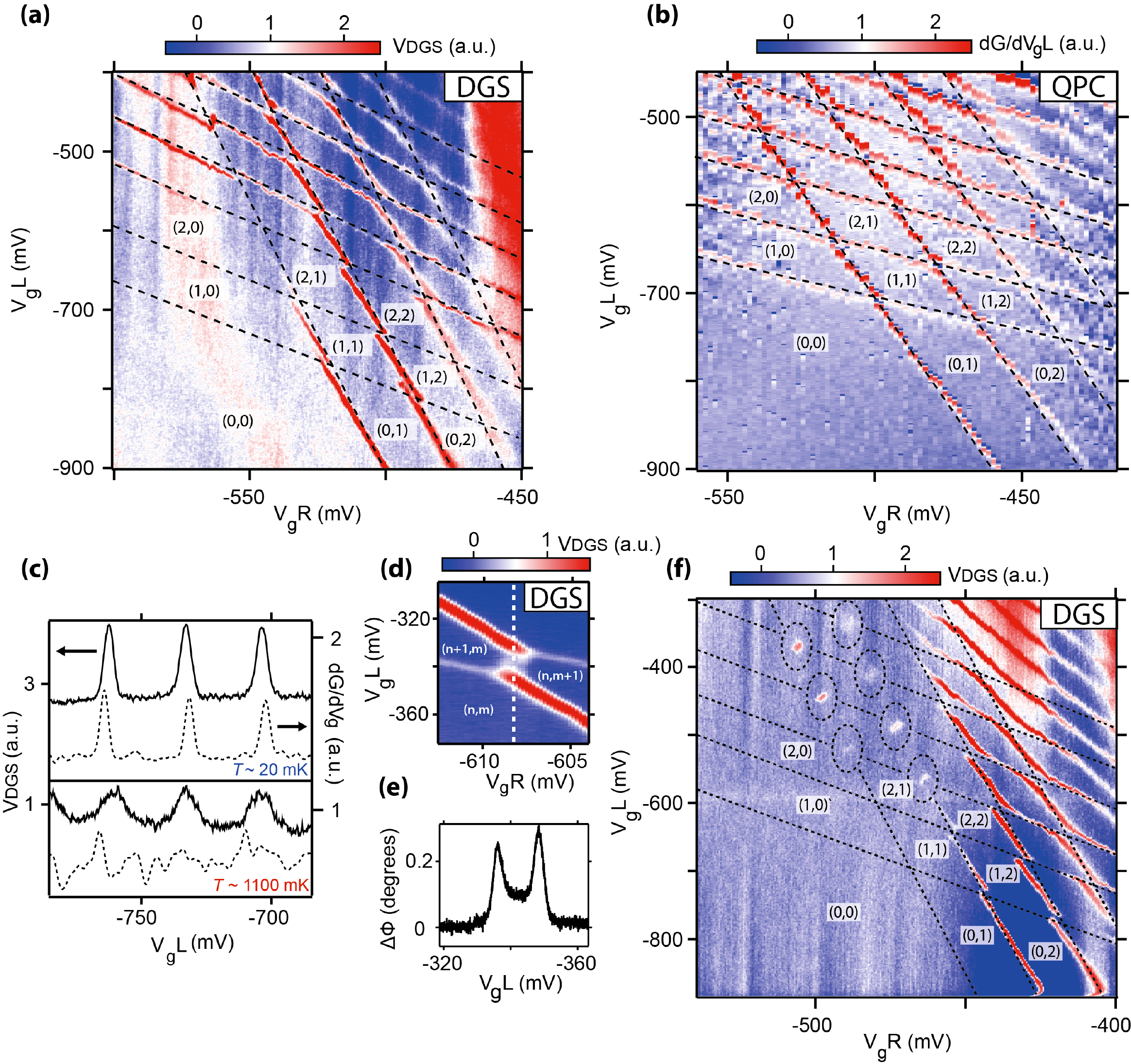}
\caption{\label{fig:photos}
\textbf{(a)} Dispersive response $V_{DGS}$ from the gate-sensor for a few-electron double quantum dot. Labels indicate number of electrons in the left and right dot. \textbf{(b)} Equivalent charge sensing signal from the QPC detector confirming the few-electron regime. \textbf{(c)} Temperature dependence of the sensing signal for both the DGS (left axis, solid line data) and QPC (right axis, dashed line data). The transitions are taken at a fixed $V_gR$ and offset vertically for clarity. Both detectors resolve clear sensing peaks at $T \sim$ 20 mK, with the QPC losing all sensitivity at elevated temperature $T \sim$ 1100 mK. \textbf{(d)} DGS signal in a zoomed-up region showing a double dot charge transition. \textbf{(e)} The calibrated phase response from the DGS for a slice through (d) with $V_gR$ held at -608 mV. \textbf{(f)} Dispersive response of the gate-sensor where tunnelling to the reservoirs is suppressed in the few-electron limit. High tunnel rate, intra-dot transitions remain visible.}
\vspace{-0.5cm}
\end{figure*}

Our device integrates a QPC charge sensor together with the DGS and allows simultaneous readout of the quantum dot system using both detectors. A comparison of the relative sensitivity of the QPC and DGS is shown in Fig. 2(a-d) where the response of each detector is measured as a function of the gate voltages $V_gL$ and $V_gR$ used to define a large, single quantum dot in the Coulomb blockade regime. The dispersive signal $V_{DGS}$ from the gate resonator is shown in Fig. 2(a,c), 
 with Fig. 2(b,d) showing the derivative of the conductance $G$ of the QPC with respect to $V_gL$. The sensitivity of both sensors is quantified by applying a small modulation voltage to a nearby gate, inducing periodic variation in conductance of the QPC or dispersive response of the DGS \cite{capmod}. We calibrate the detector signal $dG$ or $dV_{DGS}$ due to this modulation by comparing its amplitude to the signal response from a single electron transition. A measurement of the signal-to-noise ratio (SNR) in a given bandwidth yields the detector sensitivity. For the QPC we measure a typical charge sensitivity at 36 Hz of $\sim$ 3 $\times$ $10^{-3}$ e/$\sqrt{\mathrm{Hz}}$, corresponding to an integration time of 9 $\mu$s required to resolve a change of a single electron charge on the dot. The DGS method yields a similar integration time of 39 $\mu$s to resolve a single electron transition (equivalent to 6.3 $\sim$ $\times$ $10^{-3}$ e/$\sqrt{\mathrm{Hz}}$).

To determine the bandwidth of the dispersive gate-sensor the SNR of its response is measured with increasing frequency of the small modulation voltage applied to a nearby gate (Fig. 2(e)). This method gives a detection bandwidth of $\sim$ 10 MHz, limited by the $Q$-factor of the resonator, and consistent with the dependence of SNR with carrier frequency, as in Fig. 2(f). We further characterize the DGS by measuring how the height and width of the electron transition signal (see Fig. 2(c)) depends on applied resonator power, as shown in Fig. 2(g). Finally, we extract the relative geometric capacitive coupling between the sensor-gate and the quantum dot. The charging energy of the dot $E_c$ = $e^2/2C_{\Sigma}$, can be measured by using the DGS to sense Coulomb diamonds as a function of source-drain voltage across the dot $E_c$ = $eV_{sd}$, as shown in Fig. 2(h) (where $e$ is the electron charge and $C_{\Sigma}$ is the total dot capacitance). By measuring the period of CB oscillations we estimate that the gate-sensor geometric capacitance $C_g \sim$ 10 aF contributes $\sim$ 5 percent of $C_\Sigma$.   

For a single quantum dot biased at the point where electron $n$ and $n+1$ are degenerate, the quantum capacitance is given by $C_Q=(e^2/4k_B T_e)(C_g/C_\Sigma)^2$ \cite{Chorley,Glattli}, when the dot tunnel-rate is much larger than the resonator frequency ($k_B$ is the Boltzmann constant). This quantum capacitance shifts the resonance frequency by an amount $\Delta f \simeq C_Q f_0/ 2 C$,  ($C$ is the resonator capacitance). This frequency shift results in an observed phase response $\Delta \phi \simeq \alpha C_Q Q/C$, ($Q$ is the $Q$-factor of the resonator). The constant of proportionality $\alpha$ is of order unity at resonance and is related to the transmission coefficient of the resonator. For $T_e \sim$ 100 mK and $C_g/C_\Sigma\simeq0.05$ this formula gives $C_Q\simeq 9$ aF which is broadly consistent with our observed phase shifts of $\Delta \phi$ $\times $ 180$/\pi \simeq 0.2$ degrees. 

Having quantified the sensitivity of the gate-sensor, we now configure a double dot and show that this gate readout method can operate in the few-electron regime, where these devices are commonly operated as qubits. The double dot charge-stability diagram is detected using the dispersive gate-sensor as shown in Fig. 3(a), where regions of stable electron number are labeled (n,m), corresponding to the number of electrons in the left and right dots. We confirm that the double dot is indeed in the few electron regime by also detecting the charge configuration using the proximal QPC charge sensor, as shown in Fig. 3(b). 

Charge sensing using QPCs or SETs requires that the sensor be kept at a value of conductance where sensitivity is maximized. This is typically achieved by applying additional compensating voltages to gates when acquiring a charge-stability diagram. It is worth noting that gate-sensors do not require such offset charge compensation or gate voltage control. Of further practical use, we find that DGSs are robust detectors at elevated temperatures, in contrast to QPC charge sensors which suffer from a thermally broadened conductance profile and suppressed sensitivity with increasing temperature. The single-electron response of both QPC and DGS  can be compared in Fig. 3(c) for $T \sim$ 20 mK and $T \sim$ 1100 mK.

The gate-sensor can be made to detect both intra- and inter-double dot tunnelling transitions, as shown in Fig. 3(d) which depicts a close-up region of the charging diagram. A line-profile of the transitions (Fig. 3(e)) indicates that the DGS is most sensitive to electron transitions from the right reservoir, due to its position, but is capable of distinguishing all transitions. Near an intra-dot transition, the quantum capacitance for the double dot can be shown to be $C_Q^{dd}=(e^2/2 t)(C_g/C_\Sigma)^2$ where $t$ is the tunnel coupling energy of the double dot \cite{Petersson}. As for the single dot above, the phase shift (in radians) is $\Delta \phi \simeq \alpha C_Q^{dd} Q/C$. The measured phase shift $\Delta \phi \simeq 0.1$ degrees for the intra-dot transition is near half the shift for transitions to the leads, consistent with a tunnelling coupling $t/h\simeq 8$ GHz. 

Increasing the tunnel barriers between the double dot and the reservoirs suppresses the gate sensing signal when the tunnel rate drops substantially below the detector resonance frequency ($f_0 \sim$ 704 MHz). This regime is reached in Fig. 3(f), where transitions to the reservoirs are suppressed, but intra-dot transitions remain visible as these occur at a tunnel frequency above $f_0$. The observation of the intra-dot transition in the few-electron regime is important since it is this signal that forms the basis of spin qubit readout in these devices \cite{Petersson,PeterssonPetta,Hanson:2007eg}. Of further note, in contrast to QPC or SET detectors that exhibit a broadband back-action spectrum \cite{Gustavsson:2008gb}, gate-sensors act-back on the qubit at a single, adjustable frequency. 

The demonstration that $in$ $situ$ surface gates also serve as readout detectors with comparable sensitivity to QPCs is perhaps unexpected, given that the geometric gate-to-dot capacitance is only $\sim$ 5 percent of the the total capacitance. Readout using gate-sensors, however, makes use of the quantum capacitance which as we have shown, can be of the same order as the geometric contribution ($C_g \simeq C_Q$). Gate-based readout then, has potential to address the significant challenge of integrating many QPC or SET detectors into large arrays of quantum dots, for instance, in the scale-up of spin qubit devices. The use of wavelength division multiplexing techniques \cite{Stevenson,Buehler} would further allow each gate in an array to be independently and simultaneously read out at a unique frequency. Such an approach will also likely apply to systems without source-drain reservoirs altogether, such as donor qubits \cite{Morello:2010ga}, or in the readout of Majorana devices \cite{Majorana}. 

We thank X. Croot and S. D. Bartlett for technical assistance and discussions. J.M.H. acknowledges a CSIRO student scholarship and use of CSIRO facilities for Nb inductor fabrication. We acknowledge funding from the U.S. Intelligence Advanced Research Projects Activity (IARPA), through the U.S. Army Research Office and the Australian Research Council Centre of Excellence Scheme (Grant No. EQuS CE110001013).\\

$\dagger$ email: david.reilly@sydney.edu.au
\small

\begin{thebibliography}{37}
\bibitem[{\citenamefont{{A modulation in the dispersive response by a nearby
  gate can be understood as the variation in capacitance with depletion of
  electrons surrounding the DGS.}}()}]{capmod}
\bibinfo{author}{\bibnamefont{{A modulation in the dispersive response by a
  nearby gate can be understood as the variation in capacitance with depletion
  of electrons surrounding the DGS.}}}
\end{thebibliography}

\end{document}